# A computationally efficient compact model for ferroelectric FETs for the simulation of online training of neural networks

Darsen D. Lu[1], Sourav De[1], Mohammed Aftab Baig[1], Bo-Han Qiu[1] and Yao-Jen Lee[2]

[1] Institute of Microelectronics, National Cheng Kung University (NCKU), Tainan, Taiwan, R.O.C.
[2] Taiwan Semiconductor Research Institute, Hsinchu, Taiwan, R.O.C.

E-mail: darsenlu@mail.ncku.edu.tw, yjlee@narlabs.org.tw

## Abstract

Tri-gate ferroelectric FETs with $Hf_{0.5}Zr_{0.5}O_2$ gate insulator for memory and neuromorphic applications are fabricated and characterized for multi-level operation. The conductance and threshold voltage exhibit highly linear and symmetric characteristics. A compact analytical model is developed to accurately capture FET transfer characteristics, including series resistance, coulombic scattering, and vertical field dependent mobility degradation effects, as well as the evolvement of threshold voltage and mobility with ferroelectric polarization switching. The model covers both sub-threshold and strong inversion operation. Additional measurements confirm ferroelectric switching as opposed to carrier-trapping-based memory operation. The compact model is implemented in a simulation platform for online training of deep neural networks.

## 1. Introduction

Ferroelectric FETs (FeFETs) with hafnium-based dielectrics is being heavily researched as promising device for logic, memory, and neuromorphic applications ever since the proposal of negative capacitance FETs (NCFETs) for low power electronics [1] and the discovery of CMOS-compatible and highly-scalable HfZrO (HZO) ferroelectric material [2]. In particular, FeFETs are very promising for deep neural network applications given its low write latency, low cycle-to-cycle variability, good endurance, and highly symmetric and linear multi-level switching characteristics [3]. In order to apply ferroelectric FET technology to macro and system-level benchmarking [9] and design/synthesis tools [12], a compact analytical model is required. Conventional SPICE-oriented compact model can accurately model device output behavior. New compact models for ferroelectric FETs (including NCFETs) have been developed [13]. However, the simulation of on-chip neural network (NN) training requires millions of sample data to pass through hundreds of millions of synaptic devices. Full SPICE simulation with traditional compact model is simply too time consuming. For this reason, neuromorphic simulator such as [16] adopted and implemented a very simple formulation for potentiation and depression of resistive memory (RRAM). In [3], the formulation developed for RRAM was utilized for FeFET online NN training. Unfortunately, such approach fails to model FeFET-specific properties such as gate bias dependence. In this paper, we present a computationally efficient compact model for neural networks which is simple enough for evaluation during the simulation of large-scale neural network training. Yet it captures FeFET's gate bias dependence and is valid from sub-threshold to strong inversion region.

## 2. Device fabrication and measurements

Tri-gate FeFETs are fabricated on silicon-on-insulator wafer with a gate-first process [5]. The gate dielectric consists of atomic-layer-deposited $Hf_{0.5}Zr_{0.5}O_2$. Relatively thick (10nm) HZO layer is employed to ensure sufficient memory window. A 40-second rapid thermal annealing process at 700 °C after TiN gate deposition is performed for ferroelectricity.

The characterization of drain current ($I_d$) while sweeping the gate voltage ($V_{gs}$) over a wide range is convenient for compact model development. However, accurate characterization of the multi-level $I_d$-$V_{gs}$ during potentiation (programming) and depression (erasure) of FeFET is not straight-forward, since the ferroelectric state may be altered during DC $I_d$-$V_{gs}$ sweep.

For these reasons, we applied a special pulsing scheme prior to each $I_d$-$V_{gs}$ measurement (Fig. 1). A 1μs reset pulse ($V_{reset}$) ensures the FeFET returns to its erased (high threshold voltage ($V_{th}$)) state. Another 1μs program pulse ($V_{pgm}$) sets the device to the target polarization state. Subsequently $V_{gs}$ is swept from $V_{start}$ to $V_{stop}$ to measures the transfer characteristics.

The choice of $V_{start}$ is found to be very important. When $V_{start}$ lies in the sub-threshold region, the electric field is, in principle, nearly zero, and there is very little chance of ferroelectric switching. We therefore try to keep $V_{start} < V_{th}$ - 1.0V to ensure $I_d$-$V_{gs}$ covers the sub-threshold region over a 1V range and $V_{th}$ measurement is accurate. As a sanity check, for a multi-fin device with $L_g$=70nm, we apply the waveform as shown in Fig. 1 without $V_{pgm}/V_{ers}$ pulses, fix $V_{reset}$ to -5.0V and vary $V_{start}$ from -0.5V to -4.5V in steps of -0.5V, while keeping $V_{stop}$ at +4.0V. $V_{th}$ is found to be identically 0.4V, and there is negligible shift in $I_d$-$V_{gs}$ characteristics as function of $V_{start}$.

Fig. 2(a) shows measured $I_d$-$V_{gs}$ series as we vary $V_{pgm}$ in 0.2V steps. The measured device has 10 fins, fin height of 30nm, and fin width of 50nm. The curves merge together at around 4.0V because at high gate voltage, most ferroelectric domains in the multi-domain FeFET have switched, and the devices end in the same polarization state after forward $I_d$-$V_{gs}$ sweep. It can also be seen that $V_{th}$ is nearly a linear function of $V_{pgm}$ (Fig. 2(b)).

To understand the evolvement of $I_d$-$V_{gs}$ as function of $V_{pgm}$, we plotted $\Delta V_{th}$, the horizontal shift of $I_d$-$V_{gs}$ curve, using the shifting between $I_{d(5)}$ ($I_d$ after the 5$^{th}$ pulse) and $I_{d(20)}$ ($I_d$ after the 20$^{th}$ pulse) as example (Fig. 3). Surprisingly, an excellent agreement to a linear model in the on state (strong inversion) is found, which suggests that the channel inversion charge induced by ferroelectric polarization is a linear function of gate bias.

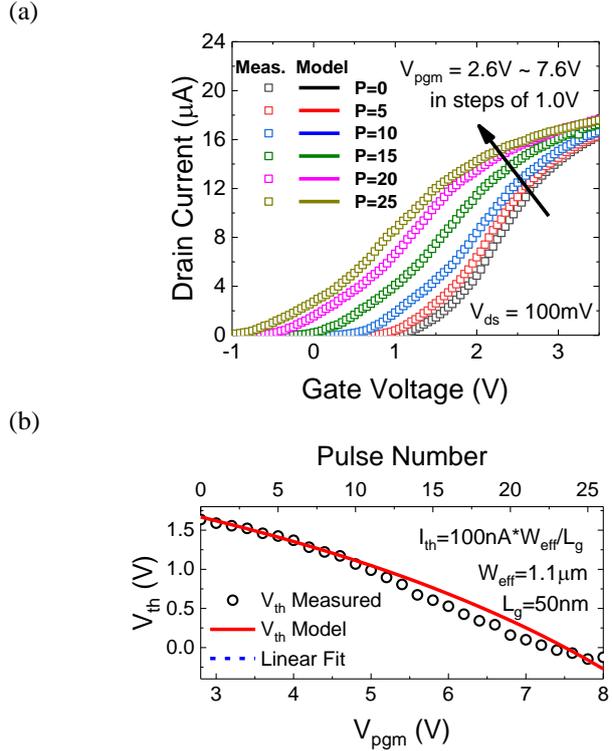

Fig. 2. (a) Measured $I_d$-$V_{gs}$ characteristics as function of $V_{pgm}$ (symbols). Only selected $I_d$-$V_{gs}$ curves (pulse number (P): 0, 5, 10, 15, 20, 25) are shown for clarity. We set $V_{start} = -V_{pgm}$ to ensure $V_{start} < V_{th} - 1.0V$ is always satisfied. Analytical compact model (lines) show good agreement with measured data (symbols). (b) Measured (symbols) and modeled (solid line) $V_{th}$ as function of $V_{pgm}$ during potentiation. Dashed line: linear fit. Data available online [19].

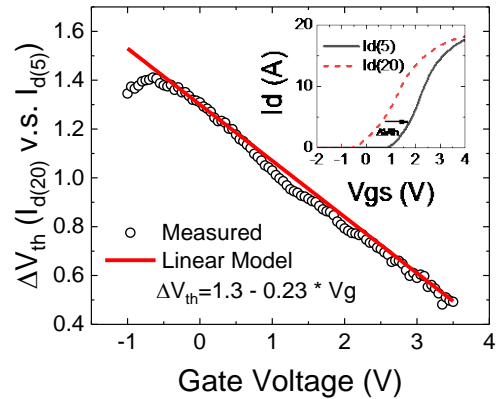

Fig. 3. Horizontal shift between $I_d$-$V_{gs}$ curves ($\Delta V_{th}$) with different $V_{pgm}$. For the entire strong inversion region, $\Delta V_{th}$ v.s. $V_{gs}$ show excellent agreement to a linear model.

## 3. Neuromorphic-oriented compact model

We aim to model FeFET memory device with a simple yet physically sound expression. The basic drain current

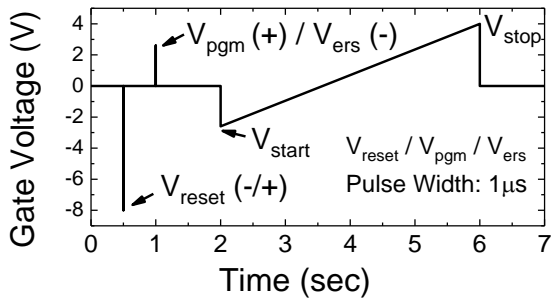

Fig. 1. Pulsing scheme for programming and erasing the FeFET devices. For programming, a reset pulse ($V_{reset}$) sets the device threshold voltage ($V_{th}$) to the highest level. This is followed by a program pulse ($V_{pgm}$) and an $I_d$-$V_{gs}$ sweep. Starting sweep voltage ($V_{start}$) is carefully chosen so that the $V_{th}$ state is not disturbed. For erase, on the other hand, $V_{reset}$ is positive and $V_{ers}$ is negative.

formulation for BSIM4 compact model [18] in the linear region as follows:

$$I_d = \frac{K \cdot V_{gsteff} \cdot V_{ds}}{1 + UA \cdot (V_{gsteff} + DELTA)^{EU} + \frac{UD}{V_{gsteff} + \frac{2kT}{q}} + RDS \cdot K \cdot V_{gsteff}} \quad (1)$$

where UA, EU and DELTA accounts for vertical field dependent mobility degradation; UD is a parameter for Coulombic scattering in moderate and weak inversion regions (mobMod=3 option in BSIM4); RDS is the source and drain series resistance. Note that the original $2 \cdot V_{th}$ term in effective vertical field calculation is absorbed into DELTA for simplicity.

In addition, to cover both weak and strong inversion regions, we adopted the effective gate overdrive function from BSIM4 [18]:

$$V_{gsteff} = \frac{\frac{nkT}{q} \ln\left[1 + exp\left(\frac{V_{geff} - V_{th}}{2\frac{nkT}{q}}\right)\right]}{\frac{1}{2} + n \cdot exp\left[-\frac{V_{geff} - V_{th} - 2 \cdot VOFF}{2\frac{nkT}{q}}\right]} \quad (2)$$

where n is the sub-threshold ideality factor, and VOFF is a parameter for $V_{th}$ shift in the weak inversion region only.

To model the linear threshold shifting, we express effective gate voltage function as:

$$V_{geff} = V_{gs} + [(FA \cdot P + FB) - (FC \cdot P - FD) \cdot V_{gs}] \quad (3)$$

where FA, FB, FC and FD are constants that control the shifting of subsequent $I_d$-$V_{gs}$ curves. The model shows good agreement with measured data over a wide range of $V_{pgm}$ (Fig. 2). Note that better fitting may be obtained with the replacement of numerator of (1) with $K \cdot V_{gsteff}^\theta \cdot V_{ds}$. However, we have chosen not to do so here to keep the model physically sound and to improve computational efficiency by reducing the number of transcendental functions in the expression. The parameter values are available in [19].

## 4. Neural network applications

Symmetric and linear potentiation / depression characteristics is crucial for online training of neural networks [9]. With properly chosen starting pulse amplitudes (+2.8V for potentiation; -0.8V for erase), excellent symmetry and linearity is achieved (Fig. 4).

In Table I, we compare this work with other FeFET technologies. Linearity of FeFET synaptic device is quantified using the non-linearity parameter, α [3][2]. In general, FeFETs show better linearity than RRAM, with lower α [9]. Note, however, that such high linearity is only possible when the program / erase voltage varies with pulse number.

In addition, unlike RRAM, FeFETs' synaptic properties are highly dependent on $V_{gs}$. The compact model presented here is very useful for evaluating figure of merits such as power consumption, circuit delay, training accuracy, etc., as function of $V_{gs}$, and ultimately used for neuromorphic circuit design.

To rule-out memory operation due to charge-trapping, double-sweep nFET $I_d$-$V_{gs}$ measurements are performed. Counter-clockwise $I_d$-$V_{gs}$ indicates ferroelectric switching (Fig. 5) [17]. We have also confirmed $V_{th}$ decrease/increase with program/erase pulse application, respectively (Fig. 5 inset).

## 5. Conclusions

Tri-gate ferroelectric FETs are characterized, showing excellent symmetry and linearity during multi-level operation. A computationally efficient and FeFET-specific compact model is developed and calibrated to measured data. The model is successfully implemented in a simulation platform for online training of deep neural networks.

TABLE I
COMPARISON OF FERROELECTRIC FET TECHNOLOGIES

| Ref. No. | Linearity ($\alpha_p$ / $\alpha_d$) | On-state resistance | # levels | Active Area (μm$^2$) |
|---|---|---|---|---|
| [3] | +1.75 / +1.46 | 559 kΩ | 32 | 1190 |
| [4] | +1.73 / +1.86 | 50 kΩ | 32 | - |
| [6] | +0.44 / -0.38 | 3.5 kΩ | 35 | - |
| [7] | +0.12 / -0.09 | 0.22 kΩ | 32 | - |
| [8] | +1.22 / -1.75 | 12.5 kΩ | 256 | 0.0034 |
| *This work* | +0.67 / -1.13 | 11.5 kΩ | 27 | 0.055 |

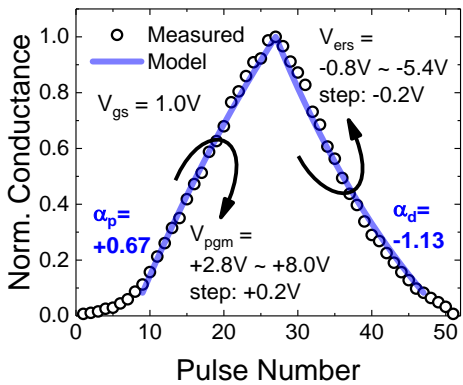

Fig. 4. Potentiation and depression of normalized channel conductance with the application of program/erase pulses. A total of 27 program pulses and 24 erase pulses are applied, which is equivalent to 4-5 bits. ($W_{eff}$=1.1μm, $L_g$=50nm) Model uses the formula in [16] for evaluating $\alpha_p$/$\alpha_p$.

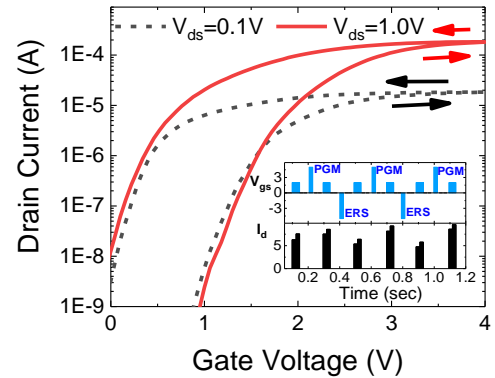

Fig. 5. DC double-sweep measurement of an $L_g$=70nm FeFET. Counter-clockwise I-V curves for nFET indicates ferroelectric switching. Inset shows device conductance with program (PGM) and erase (ERS) pulse application (For clarity, only $I_d$ during 2V read pulse is shown).

## Acknowledgements

This work was jointly supported by the Ministry of Science and Technology (Taiwan) grant MOST-108-2634-F-006-08 and is part of research work by MOST's AI Biomedical Research Center. We are grateful to Taiwan Semiconductor Research Institute for nanofabrication facilities and services, and Dr. Wen-Jay Lee and Nan-Yow Chen of National Center for High-Performance Computing for helpful suggestions on AI computation.